\documentclass[referee]{raa}            
\twocolumn
\usepackage{graphicx,times}             
\usepackage{natbib}
\usepackage{amssymb,amsmath}
\usepackage[utf8]{inputenc}
\bibpunct{(}{)}{;}{a}{}{,}

\usepackage[pagebackref=true]{hyperref}

\begin{document}

  \title{Study of Four nulling pulsars with FAST}

   \volnopage{Vol.0 (20xx) No.0, 000--000}      
   \setcounter{page}{1}          

   \author{Jingbo Wang 
      \inst{1}
   \and Jintao Xie
      \inst{2}
    \and Jing Zou
      \inst{3,4,1}
    \and Jianfei Tang
      \inst{1}
   }

   \institute{Institute of Optoelectronic Technology, Lishui University, Lishui 323000, China; {\it tjf1027@163.com}\\
        \and
             School of Computer Science and Engineering, Sichuan University of Science and Engineering, Yibin 644000, China\\
        \and
             Xinjiang Astronomical Observatory, Chinese Academy of Sciences, Urumqi, Xinjiang 830011, China\\
        \and
             University of Chinese Academy of Sciences, Beijing 100049, China\\
\vs\no
   {\small Received 2025 month day; accepted 2025 month day}}

\abstract{ We present an analysis of 4 nulling pulsars with the Five-hundred-meter Aperture Spherical radio Telescope (FAST).
For PSR J1649+2533, our results suggest mode changing rather than subpulse drifting as previously reported at lower frequencies. For PSR J1752+2359, we confirm its quasi-periodic switching between distinct emission states, but further show that the so-called ``quasi-null'' or ``RRAT-like'' state actually consists of persistent low-level emission superposed with occasional bright pulses. For PSR J1819+1305, our data confirm the modulation reported earlier, while additional weaker features are also seen. 
For PSR J1916+1023, we detect both nulling and subpulse drifting, but find no clear evidence of direct interaction between them.
These results provide new insights into the diverse manifestations of pulsar nulling, highlight the capability of FAST to detect subtle emission states, and add to the growing body of work on pulsar emission variability.
}

   \authorrunning{J. -B. Wang, J. -T. Xi,  J. Zou,  et. al }            
   \titlerunning{Study of four nulling pulsars with FAST }  
    
   \maketitle

\section{Introduction}           
\label{sect:intro}

Pulsars are commonly regarded as fast-spinning neutron stars that emit highly directional and coherent electromagnetic radiation from their poles in a steady and predictable way. According to this model, Earth observers detect a pulse precisely when the pulsar's radiation beam crosses their line-of-sight (LOS), which happens once per rotational cycle. In fact, soon after the discovery of pulsars it was found that the pulsar emission is seldom completely stable. Their pulsed emission varies from pulse to pulse in intensity, shape and phase. 

Absence of pulsed emission for several pulsar rotations was first noted by \citet{1970Natur.228...42B}. This so called pulse nulling phenomenon which has been seen in more than 100 pulsars \citep{2007MNRAS.377.1383W,2012MNRAS.423.1351B}
, and is relatively common in long period pulsars \citep{1986ApJ...301..901R}.
Heterogeneous null durations manifest at both inter-pulsar and intra-pulsar scales. The nulling fraction (NF), defined as the proportion of pulses showing no detectable emission, exhibits values from $\sim$ 0\% to $>$ 95\%. PSRs B2021+51 and B0835$-$41 demonstrate nulling behavior limited mostly to single pulses \citep{2012MNRAS.424.1197G}.
In contrast, PSR B0826$-$34 displays a fundamentally different pattern, remaining in null states during most observations \citep{1979MNRAS.186P..39D}.

There has been no satisfactory explanation for the pulse nulling since its discovery. It can be caused by a number of reasons. Given that pulsar radiation arises from phase-locked plasma oscillations, brief null states are modeled as nonlinear perturbations that episodically interfere with the necessary phase coherence. The nulling phenomenon in pulsars exhibiting drifting subpulses may stem from the geometrical configuration of a spark carousel in drift motion \citep{2008MNRAS.385.1923R}.

In contrast, pulsars exhibiting cessation of emission over extended timescales (days to years) are typically classified as intermittent pulsars - a regime representing the most extreme manifestation of nulling behavior. Recent studies show that the pulsar’s spin-down rate is significantly reduced when the intermittent pulsar in its ``off'' state \citep{2006Sci...312..549K,2012ApJ...758..141L,Camilo12,Lyne17,2020ApJ...897....8W}. 
The currents and particle acceleration powering radio emission of intermittent pulsars could stop, or the radiation beam shift away the line of sight owing to unsuitable emission geometry \citep{2010MNRAS.408.2092T}. 
Some pulsars are also known to switch between multiple distinct profile states which is called as mode changing. \citet{2007MNRAS.377.1383W,2010Sci...329..408L} 
argued that nulling and mode changing are related phenomena and differ only in the magnitude of the changes in the magnetospheric current flows.

In recent years, substantial progress has been made in understanding pulsar nulling and related emission phenomena through both large-sample statistical analyses and high-sensitivity case studies. \citet{2020AA...644A..73W} observed 20 bright pulsars at 2250 MHz with the Jiamusi 66 m telescope for unprecedented durations, discovering three new nulling pulsars and showing that null and emission sequences in most sources deviate from purely random processes, often exhibiting quasi-periodic or low-frequency modulations. They also found correlations between nulling parameters and spin-down properties, and concluded from a combined sample of 146 nulling pulsars that large nulling fractions are more closely related to spin period than to characteristic age or energy loss rate. 
\citet{2020MNRAS.408..906} compiled an extensive sample of \(\sim\)70 pulsars exhibiting periodic modulations—including periodic nulling and amplitude modulation—demonstrating that such behaviour can occur across the entire pulse profile and is physically distinct from subpulse drifting. 
\citet{2023APJ.948..32} applied a robust mixture-model method to measure nulling fractions for 22 recently discovered pulsars, eliminating biases inherent in traditional Ritchings-type algorithms, and found that previously reported correlations between NF and spin parameters may be significantly affected by such biases. 
\citet{2023MNRAS.520..4562} reported new cases of frequency-dependent nulling and weak emission during nulls, further highlighting the complexity of emission-state transitions. FAST has also been used to study individual nulling pulsars in detail 
\citep[e.g.,][]{2021RAA.21..240,2024MNRAS.528..6388} although large-sample nulling analyses with FAST are not yet available. In the context of other intermittent neutron star populations, 
\citet{2023RAA.23..104001} used FAST in the Galactic Plane Pulsar Snapshot survey to discover 76 new rotating radio transients (RRATs) and detect 48 previously known ones, revealing that many RRATs are in fact extremely nulling pulsars or weak pulsars with occasional strong pulses, suggesting a physical continuity between these populations. These recent works underscore the importance of high-sensitivity, single-pulse analyses for probing emission–null transitions, motivating the present FAST study of four nulling pulsars.
Nulling pulsars have received concentrated study through discovery and observation efforts, which may provide key evidence to explain the cause of variability in pulsar emission. Using the Five-hundred-meter Aperture Spherical radio Telescope (FAST), we have observed nulling phenomenon in 4 pulsars.

The first of our pulsars, PSR J1649+2533 is a 1015 ms source which was  in Arecibo high galactic latitude pulsar survey \citep{1995ApJ...454..826F}
. Nulling phenomenon of the source was initially identified by \citet{2004ApJ...600..905L} 
and further analyzed by \citet{2009MNRAS.393.1391H}
. It was found to have a null fraction of $\sim$ 25\%. Drifting subpulses were also identified in this pulsar.

PSR J1752+2359, the second source, is a 409 ms pulsar which was also identified in Arecibo high galactic latitude pulsar search \citep{1995ApJ...454..826F}
. Follow-up observations demonstrated that the source undergoes pulse nulling for $\sim$ 80\% of the time \citep{2004ApJ...600..905L,2014ApJ...797...18G}
. This pulsar exhibits remarkable single pulse characteristics, with bursts of reaching 100 pulses alternating with nulls of approximately 500 pulses. Our research additionally found an exponential attenuation in the pulse energy during a burst, a signature documented in only limited number of pulsars \citep{2008MNRAS.385.1923R,2010MNRAS.408..407B}
 
PSR J1819+1305, the third one, is a 1060 ms source which was identified by Swinburne intermediate-latitude pulsar survey \citep{2001MNRAS.326..358E}
. This pulsar undergoes pulse nulling roughly half of the time. The observed null intervals of the pulsar exhibited a strong periodicity of $\sim$ 57 times the period \citep{2008MNRAS.385.1923R}.
 
PSR J1916+1023, the last pulsar, is a 618 ms source which was discovered by the Parkes multibeam pulsar survey \citep{2004MNRAS.352.1439H}
. Short bursts of emission are observed from this pulsar, which appears to spend about half of the time in a null state.

In the following section, we review our observation programme used to collect our data. Our results are presented in Section 3. Lastly, a discussing of our results and conclusions are given in Section 4.




\section{Observations}
\label{sect:Obs}

The data presented in this study were acquired using the
FAST. FAST is a Chinese national major scientific infrastructure managed by the National Astronomical Observatories, Chinese Academy of Sciences \citep{2020RAA....20...64J}
. FAST has a very large collecting area allowing us to observe single pulses with high S/N for all the four pulsars. The observations of the four pulsars were performed on 2019 June and July (MJD 58663, MJD 58659, MJD 58660 and MJD 58684 for PSRs J1649+2533, J1752+2359, J1819+1305 and J1916+1023, respectively). 1 hour of observations for each pulsar were recorded using the central beam of the 19 beam receiver covering from 1050 MHz to 1450 MHz. The entire quantity of pulses are 3546, 8803, 3715 and 2225 for PSRs J1649+2533, J1752+2359, J1819+1305 and J1916+1023, respectively.
No polarization calibration observations were performed for all the four pulsars.

The data were captured with a digital backend based on Reconfigurable Open Architecture Computing Hardware generation 2 (ROACH 2). The captured data were recorded in search mode PSRFITS format \citep{2004PASA...21..302H} 
with a time resolution of 49.152 $\mu$s and frequency channel bandwidth of 0.122 MHz, respectively. Individual pulses are extracted with 1024 phase bins per pulse period using the {\sc dspsr} software package \citep{2011PASA...28....1V}. 
The ephemeris of the four pulsars were obtained from the ATNF pulsar catalogue \citep{2005AJ....129.1993M}.

We mitigated narrow band and aliased signals radio-frequency interference by excising channels within 5\% of the band edge and those with a level substantially above a median-smoothed bandpass using the {\sc psrchive} program {\sc paz} \citep{2004PASA...21..302H} 
, respectively. {\sc psrsalsa} packages were used to investigate the single pulse modulation in detail.

\section{Results}
\subsection{PSR J1649+2533}
 Figure~\ref{fg:1649stack} shows the pulse-stack of PSR J1649+2533 obtained using the FAST on 2019 20th July. This pulse-stack and profile has been obtained after summing in frequency. As shown in Figure 1, the average pulse profile consists of only one main component and it is not symmetric. The profile is asymmetric, with a broader left side that may indicate an additional component. The pulse profile at 430 MHz is very similar to our observations at a centre frequency of 1250 MHz. The pulse widths at 50\% (W\(_{50}\)) and 10\% (W\(_{10}\)) of the peak intensity were measured  from the integrated pulse profiles.
 The uncertainties for W\(_{50}\) and W\(_{10}\) were estimated based on the off-pulse baseline noise. The uncertainty is calculated as half the difference between the widths measured at the nominal threshold ( 50\(\%\) and 10\(\%\) of the peak) shifted up and down by one root-mean-square (RMS) of the baseline. The W\(_{50}\) and W\(_{10}\) at 1250 MHz are \(24.72 \pm 0.05\) and \(37.16\pm 0.11\) milliseconds, respectively; which is almost identical to that measured at 430 MHz \citep{2004ApJ...600..905L}.

As shown in Figure~\ref{fg:1649mode} and the left panel of Figure~\ref{fg:1649stack}, the pulsar exhibits mode changing behavior with different emission phases. The two modes are discriminated by the peak emission phase and pulse width. 
Mode A constitutes the statistically dominant state, whereas Mode B occurs sporadically. The average profile of mode A is wider than that of mode B. The W$_{50}$ of mode A and mode B are 9.5$^{\circ}$ and 7.0$^{\circ}$, respectively. The peak emission phase of mode A is 4.25$^{\circ}$ earlier in longitude than that of mode B. The duration of both mode A and mode B are short and in the range of 1 to 30 pulse periods with peak at about 3 pulse periods. As seen from Figure~\ref{fg:1649mode_dist}, the pulsar switches from one mode to another frequently. 

\begin{figure}[h]
    \centering
    \includegraphics[width=0.45\textwidth]{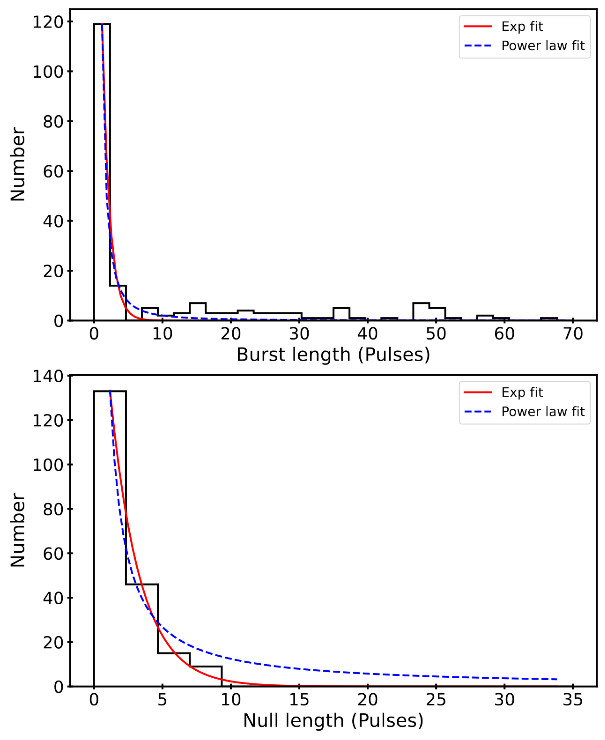}
    \caption{Burst and null duration distributions for PSR J1649+2533.}
    \label{fg:1649hist}
\end{figure}

\begin{figure}[h]
    \centering
    \includegraphics[width=0.45\textwidth]{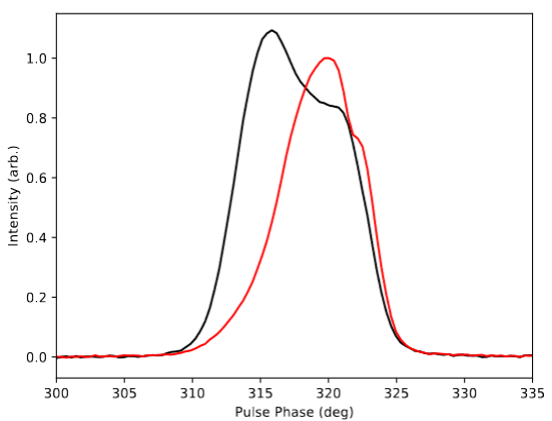}
    \caption{The average profiles of the mode A (black line) and mode B (red line) of PSR J1649+2533.
    }
    \label{fg:1649mode}
\end{figure}

\begin{figure}[h]
    \centering
    \includegraphics[width=0.45\textwidth]{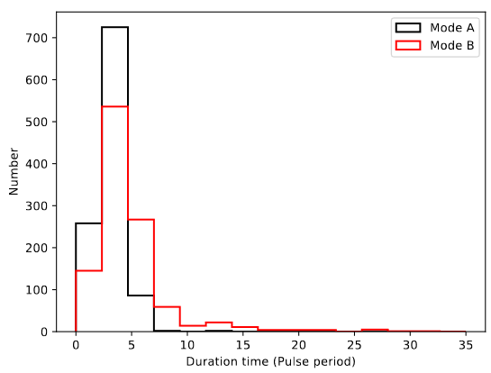}
    \caption{The duration of the mode A (black line) and mode B (red line) of PSR J1649+2533.}
    \label{fg:1649mode_dist}
\end{figure}

The single-pulse energy was calculated by integrating under the pulse profile following baseline subtraction, while reference off-pulse values were obtained by averaging equivalent-duration segments from quiet periods. Adopting methodology analogous to \citet{2010MNRAS.408..407B}
, we identify individual null pulses, where pulses with intensity smaller than 5 $\sigma_{ep}$ are categorized as nulls, where $\sigma_{ep}$ is the uncertainty of the pulse energy computed from the standard deviation energy in the off-pulse window. Following the procedure presented by \citet{1976MNRAS.176..249}
, we progressively subtracted portions of the off-pulse energy histogram from the on-pulse histogram until the summed counts in bins with E $\leq$ 0 reached zero. The fraction obtained is the null fraction (NF), indicating the proportion of null pulses in the sequence. The uncertainty of NF is given by $\sqrt{n_{\mathrm{p}}}/N$, where $n_{\mathrm{p}}$ denotes null pulse counts and N is the total pulse number. Null and burst lengths were determined as the timespan between the first pulse energy below (above) the threshold mentioned above to the next above (below) the threshold.

 Our observations indicate a NF of 20.6\% $\pm$ 0.8\% which is a bit lower than previous results \citep{2004ApJ...600..905L,2009MNRAS.393.1391H}
 . Note that the previous single pulses observations were conducted at 430 MHz and 327 MHz. 
 
\begin{figure}[h]
    \centering
    \includegraphics[width=0.5\textwidth]{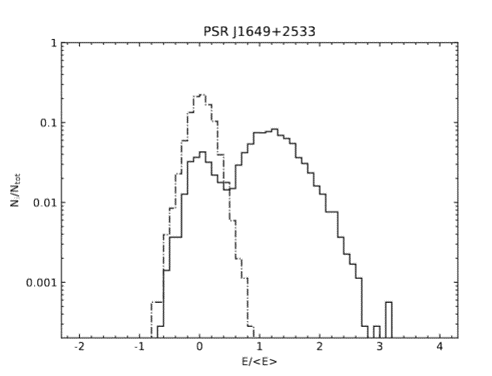}
    \caption{
    Normalized pulse energy distributions for PSR J1649+2533: on-pulse (solid line) vs. off-pulse (dashed line).}
    \label{fg:1649ped}
\end{figure}

 As shown in the right panel of Figure~\ref{fg:1649stack} 
 , the pulse stack is dominated by short bursts and nulls. The average length of the burst is 31.3 pulses with standard deviation of 28.7 pulses. The average length of the null is 9.1 pulses with standard deviation of 6.9 pulses. 
The null and burst length statistic are shown in Figure~\ref{fg:1649hist}. To test whether the null and burst durations are consistent with a stochastic process, we modeled their distributions with exponential and power-law functions.The null length are well fitted by an exponential distribution and the burst-length distribution is better fitted by a power-law.
 The difference in the intensity of the pulses in the on and null states could be more precisely characterized by investigating the pulse energy distribution. 
 Single-pulse energies were quantified by longitudinal integration of on-pulse intensities. The pulse energy distribution displays in Figure~\ref{fg:1649ped} along with the pulse energy distribution derived by integrating off pulse bins using an identical pulse longitude interval. The FAST observations lacked flux calibration, therefore, the obtained pulse energies were scaled by the average pulse energy $<E>$.
 The pulse energy distribution shows unambiguous evidence supporting the existence of bi-modality. 
 Such bi-modality might naturally be anticipated considering the distribution comprises a combination of burst and null pulses that ought to exhibit distinct mean brightness characteristics.
 Stochastic noise variations can be expected to increase the width of the fundamental pulse energy distributions, this effect being quantified through the off-pulse distribution illustrated by the dotted histogram in Figure~\ref{fg:1649ped}. That may be the reason why the distribution is continuous between the pulses and nulls. As shown in Figure~\ref{fg:1649ped}, the pulse energy of the bursts fluctuate in a small range and the observations at 327 MHz exhibit similar characteristic \citep{2009MNRAS.393.1391H}.


\subsection{PSR J1752+2359}
\begin{figure}[h]
    \centering
    \includegraphics[width=0.45\textwidth]{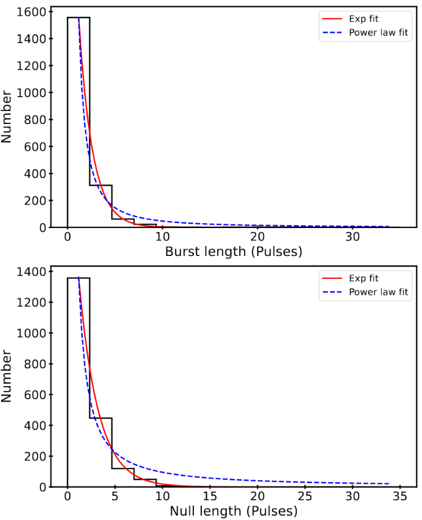}
    \caption{Burst and null duration distributions for PSR J1752+2359.}
    \label{fg:1752hist}
\end{figure}

\begin{figure}[h]
    \centering
    \includegraphics[width=0.5\textwidth]{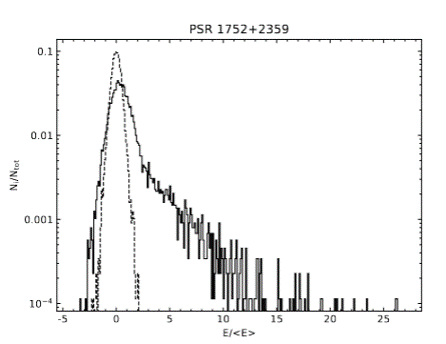}
    \caption{
    Normalized pulse energy distributions for PSR J1752+2359: on-pulse (solid line) vs. off-pulse (dashed line).}
    \label{fg:1752ped}
\end{figure}

The single-pulse sequences and mean profile of PSR J1752+2359 are shown in Figure~{\ref{fg:1752stack}}.
The overall mean profile has at least three components and is similar to that at 430 MHz \citep{1995ApJ...454..826F}
. The  W$_{50}$ and W$_{10}$ at 1250 MHz are \(3.03\pm 0.03\) and \(18.46\pm 0.29\) millisecond which are significantly wider than those at 430 MHz \citep{2004ApJ...600..905L}
. This is uncommon since pulse profile are usually narrower at high frequency than that at low frequency.
The pulsar exhibits interesting single pulse behaviour with NF of 83\% $\pm$ 1\% and spending most of the time in a
``quasi-null'' state. For PSR J1752+2359, we define the ``quasi-null'' state as an emission state in which the pulsar appears to be in a RRAT-like mode with only sporadic detectable strong pulses superposed on long null sequence.
In contrast, a traditional nulling state refers to intervals in which no detectable emission is present even after integrations,consistent with either a temporary cessation of radio emission or emission remaining below the detection sensitivity of the observation.
The separations between ``on-states'' range from 217 to 897 pulses with an average of 537 pulses. The duration of the ``on-states'' ranges from 57 to 204 pulses with an average of 89 pulses. The burst lengths histograms of PSR J1752+2359 shown in Figure~\ref{fg:1752hist} peaks at a few pulses which indicates the ``on-states'' are mixed with short nulls.
As shown in Figure~\ref{fg:1752hist}, there is a large number of isolated burst pulses, which occur in the ``quasi-null'' state. They appear to be random during the ``quasi-null'' state, and throughout the entire emission of the pulsar. 
We modeled the burst and null length distributions using both power-law and exponential forms and applied the KS test. For PSR J1752+2359, the null-length distribution is more consistent with an exponential law, whereas the burst length the burst-length distribution is better fitted by an power-law.
The on-pulse and off-pulse energy histograms are shown in Figure.~\ref{fg:1752ped}. 
The single energy distribution shows that a number of pulses have energies exceeding 10 times the mean pulse energy. These events are indicative of strong, and potentially giant, pulses based on conventional giant-pulse criteria. Note that a high energy tail is obviously seen in our observations. The total number of giant pulses is 186 giving a giant pulse rate about one pulse in 47 pulses. The energy of the strongest giant pulse exceeds the energy of the average pulse by a factor of 26. 

\begin{figure}[h]
    \centering
    \includegraphics[width=0.5\textwidth]{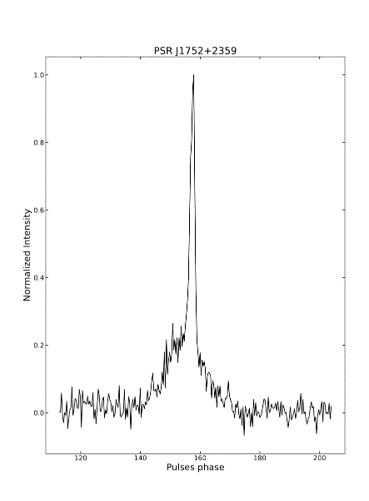}
    \caption{Integrated pulse profile over all the nulls for PSR J1752+2359.}
    \label{fg:1752offpro}
\end{figure}

The Integrated profile of all the nulls was formed and is shown in Figure~\ref{fg:1752offpro}. 
The integrated profile of all the nulls clearly suggests there is low-level emission for the pulsar. Compared with the overall mean profile, the integrated profile of all the nulls is obviously thinner.
In previous studies \citep[e.g.,][]{2021RAA.21..240}
, this emission state was classified as a ``quasi-null'' state or ``RRAT-like'' emission, characterized by the apparent absence of regular pulses except for sporadic strong ones. Our analysis reveals that, after removing these sporadic bright pulses and integrating the remaining pulses, a clear cumulative profile is still detected. This indicates the presence of persistent low-level emission throughout this state, implying that it is in fact a combination of a weak-emission mode and sporadic bright pulses rather than a true quasi-null or purely RRAT-like state. Such a hybrid state appears to be rare among known pulsars and may provide new insights into the relationship between nulling, mode changing, and sporadic strong-pulse emission.

\subsection{PSR J1819+1305}
The pulse stack of PSR J1819+1305 is shown in Figure~\ref{fg:1819gray}. Its averaged pulse profile is asymmetric and consists of at least four components. The observation at 327 MHz also exhibits asymmetric three-component profile but the three-components are more  separated and the middle component is the strongest \citep{2008MNRAS.385.1923R}. 
As shown in Figure.~\ref{fg:1819gray}, there is strong intensity modulation due to nulling with emission absent for about 33\% $\pm$ 1\%  of the time, which is a bit lower than that at 327 MHz. 

\begin{figure}[h]
    \centering
    \includegraphics[width=0.5\textwidth]{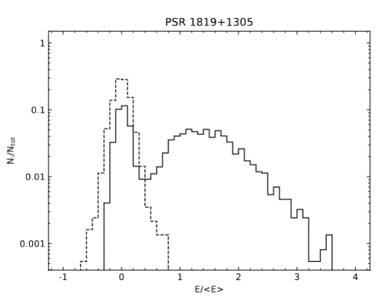}
    \caption{
    Normalized pulse energy distributions for PSR J1819+1305: on-pulse (solid line) vs. off-pulse (dashed line).}
    \label{fg:1819en}
\end{figure}
\begin{figure}[h]
    \centering
    \includegraphics[width=0.45\textwidth]{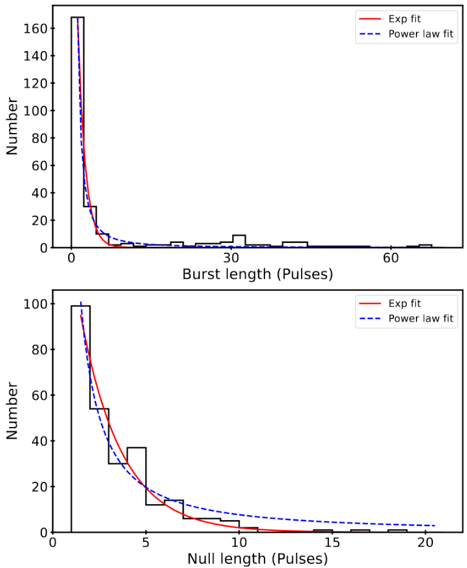}
    \caption{
    Burst and null duration distributions for PSR J1819+1305.}
    \label{fg:1819lnull}
\end{figure}

The pulse energy histogram in Figure~\ref{fg:1819en} shows bimodel distribution, which is similar with that at 327 MHz. The peak around zero demonstrates that there is a identifiable population of nulls and they are indistinguishable from weak normal pulses.
The distributions of the length of burst and null are presented in Figure~\ref{fg:1819lnull}. The null-length distribution apparently is dominated by short nulls. 
The burst-length distribution shows a similar behavior, being dominated by short bursts, though a small fraction of longer bursts is also present. We fitted both the null-length and burst-length distributions with exponential and power-law models.For PSR J1819+1305, the burst-length distribution is better fitted by an power-law, while the null-length distribution is consistent with a exponential model.

The burst-length distribution is obviously dominated by the bursts with a few and a few tens of periods. 
Significantly, some extended emission episodes are observed – two over about 140 and four between 60 and 90 periods – demonstrating that the observed periodicity of the null bunches is not a strict cadence. The null-length distribution is statistically predominated by short nulls with a few period. However, as with the burst histogram, note that a very long null sequences are also observed. The average burst and null lengths are 33 and 17 pulses, respectively, which is much longer than that at 430 MHz. 

\begin{figure}[h]
    \centering
    \includegraphics[width=0.35\textwidth, angle=-90]{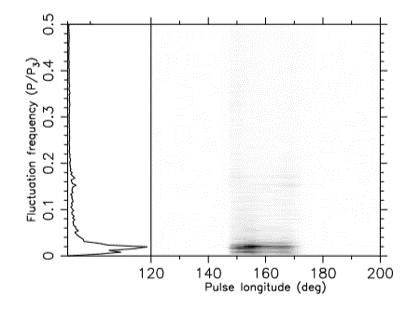}
    \caption{Fluctuation analysis of the emission of PSR J1819+1305.}
    \label{fg:1819lrfs}
\end{figure}

As reported by \citet{2008MNRAS.385.1923R} and \citet{2003ApJ...594..943N}
, PSR J1819+1305 exhibits ``periodic nulls'' and Figure.~\ref{fg:1819gray} illustrates the interesting phenomenon. Irregularities in the ``null'' spacings can be easily found and at multiple instances the radiation persists for more than 100 pulses. The pulse sequences exhibits pronounced null-state modulation exhibiting an approximate 50-pulse cyclicity which is just the sum of the average burst and null lengths mentioned above. \citet{2008MNRAS.385.1923R} 
reported that at 327 MHz PSR J1819+1305 exhibits an asymmetric three-component profile, with the central (core-like) component being the strongest and the leading and trailing components being relatively weaker and more clearly separated. In our FAST 1250 MHz observation, the average profile also shows three components with an overall width comparable to that at 327 MHz, but the components are less clearly separated, and the leading component is now the most prominent. The longitude-resolved fluctuation spectrum (LRFS) shown in Figure.\ref{fg:1819lrfs} indicates all the three pulse components demonstrate a dominant oscillatory pattern at approximately 0.021$\pm$0.002 cpp (cycle per period), corresponding to a period about 50 pulses, consistent with the modulation reported by \citet{2008MNRAS.385.1923R}. 
Features corresponding to P$_3$ of 0.16$\pm$ 0.02 cpp are only prominent in the outer component which is also consistent with results  reported by \citet{2008MNRAS.385.1923R}. 
We find no evidence for P3 $\approx$ 3 P1 modulation as reported in \citet{2008MNRAS.385.1923R}. 
No subpulse drifting was found in the pulsar.

\subsection{PSR J1916+1023}
\begin{figure}[h]
    \centering
    \includegraphics[width=0.45\textwidth]{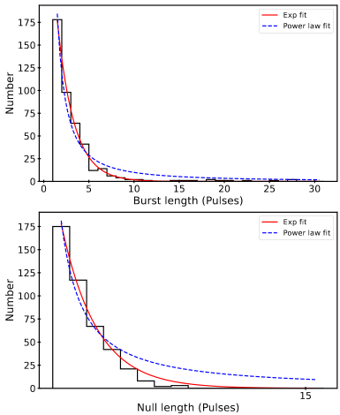}
    \caption{
    Burst and null duration distributions for PSR J1916+1023 obtained from the FAST observations.}
    \label{fg:1916lnull}
\end{figure}

As shown in Figure~\ref{fg:1916Intergrate}, PSR J1916+1023 has at least two components and there is bridge emission between the leading and trailing component.
The pulsar stays in a null state for 84\% $\pm$ 2\% of the period exhibiting numerous brief emission bursts. The nulling fraction for the pulsar is similar to that of PSR J1752+2359. 
The burst lengths histograms of PSR J1916+1023 shown in Figure~\ref{fg:1916lnull} peaks at a few pulses which indicates the “on-states” are mixed with short nulls like J1752+1359. 
Both the burst- and null-length distributions are modeled using exponential and power-law.For PSR J1916+1023, the burst-length distribution is better fitted by an exponential law, while the null-length distribution is more consistent with a power-law.
These short nulls could be also easily spotted in Figure~\ref{fg:1916gray}. Unlike PSR J1752+2359, there is almost no isolated burst between those long nulls. The sub-pulse drifting for both the leading and trailing components could be clearly seen from Figure~\ref{fg:1916gray}. Due to the high nulling fraction, the sub-pulse drifiting is frequently interrupted by both short and long nulls. 444P$_2$ and P$_3$ basically remain unchanged and are about 10 degree and 12P, respectively. In Figure~\ref{fg:1916gray}, it is clear to see that PSR J1916+1023 shows a good match of the expected driftband slope during the nulling state between about 2945 to 2085 pulses, which suggests that it may continue to drift during nulls.
The pulse energy histogram in Figure~\ref{fg:1916en} shows bi-model distribution and the pulses and nulls can be discriminated. Note that there are a few pulses with energy larger 10 $<E>$, with the strongest reaching only  11 $<E>$, While this marginally satisfies the commonly used energy-based threshold for giant pulses, the small excess over the threshold,suggests that these events are more appropriately described as strong pulses.


\begin{figure}[h]
    \centering
    \includegraphics[width=0.45\textwidth]{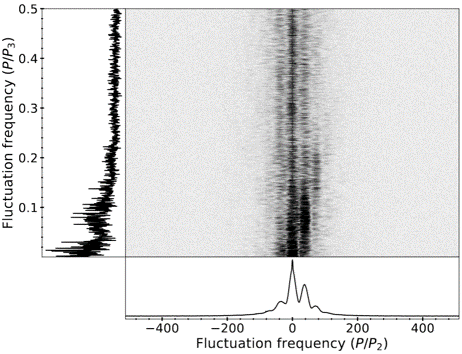}
    \caption{Two-dimensional fluctuation spectrum of PSR J1916+1023.
    }
    \label{fg:1916ldist}
\end{figure}

\begin{figure}[h]
    \centering
    \includegraphics[width=0.5\textwidth]{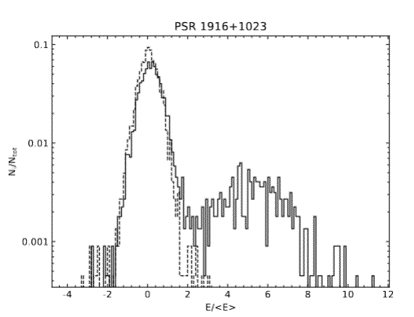}
    \caption{
    Normalized pulse energy distributions for PSR J1916+1023: on-pulse (solid line) vs. off-pulse (dashed line).}
    \label{fg:1916en} 
\end{figure}

\section{Discussion and Summary}

PSR J1649+2533 was identified as drifting subpulse pulsars by \citet{2004ApJ...600..905L} and \citet{2009MNRAS.393.1391H}.  
PSR J1649+2533 shows clear evidence of nulling in our FAST observations. However, unlike previous low-frequency studies which reported the presence of subpulse drifting , our FAST data at 1250 MHz do not show any drifting features in the fluctuation spectra. Instead, we find strong evidence of mode changing: the pulsar switches between distinct emission modes with noticeably different average profiles and fluctuation properties. This indicates that the emission behavior of PSR J1649+2533 is possibly frequency dependent, with subpulse drifting appearing at lower frequencies while mode changing dominates at higher frequencies. In addition, the nulling fraction, pulse energy distribution and pulse profile are all similar at both frequencies. 


The burst pulses in some nulling pulsars are strong enough to create a bimodal intensity distribution \citep{2012MNRAS.424.1197G}. 
PSR J1752+2359 does not show such a bimodal distribution because of the presence of many weak energy pulses.
Weak emission has been observed during phases classified as null \citep{2005MNRAS.356...59E} 
and might raise up doubts whether weak pulses are sometimes misidentified ``true'' nulls. Therefore, we investigated this by adding up all the  nulls together.
The profile obtained from these nulls are shown in Figure~\ref{fg:1752offpro}.

Weak emission throughout the null is clearly seen for J1752+2359. No such weak emission is detected for a much longer observation with GMRT at 320 MHz \citep{2014ApJ...797...18G} 
which indicate it may be frequency dependent. We also add up all the  nulls for the other three pulsars, no weak emission is seen. Such weak emission may be detected in more pulsars by FSAT in the future. 
\citet{2021RAA.21..240} observed PSR J1752+2359 for 1.5 hours with FAST at 1250 MHz and reported a nulling fraction of about 80$\%$. They showed that the pulsar exhibits two distinct emission states: a normal state with continuous pulse emission and a RRAT-like state with sporadic emission, with a quasi-periodic switching between the two states on a timescale of $\sim$568 rotation periods. Our one-hour FAST observation yields a slightly higher nulling fraction of 83$\%$, but confirms the existence of the same two emission states. While 
\citet{2021RAA.21..240} quoted $\sim$568 periods as the average switching timescale, inspection of their Figure X shows that the actual durations vary considerably, a trend also seen in our data. We measure a broader range of cycle lengths, from just over 200 pulses to nearly 1000 pulses. Importantly, in the RRAT-like state, after removing the sporadic strong pulses and integrating the remaining pulses, we detect a clear cumulative pulse profile, indicating the presence of a persistent low-level emission mode rather than complete radio silence. 
\citet{2021RAA.21..240} also presented polarization properties for the two states, whereas our data were not polarization-calibrated and therefore do not allow a similar analysis. These differences, together with the consistency in the overall phenomenology, provide complementary constraints on the emission-state switching mechanism in this source.

At the frequency of 111 MHz, the pulsar PSR J1752+2359 has been found to produce giant pulses. The energy of the strongest giant pulses was measured as two hundred times the average pulse energy. Our observations also shows giant pulses up to 32 times the mean pulse energy. More energetic giant may be detected with longer observation. 
In fact, An increasing number of pulsars with giant pulse have been discovered recent years, which indicate giant pulse may be more common than we thought before. 

Previous observations for PSRs J1649+2533, J1752+2359 and J1819+1305 were carried out at lower frequency around 430 MHz. The NF we obtained is different with that at 430 MHz for the three pulsars. The duration of our observations was 1 hour for each pulsar. However, such brief observational durations, particularly for pulsars with approximately 1-second rotational periods, do not provide adequate nulls for a robust comparison of their nulling fraction. significantly longer observations are needed to acquire a substantial sample of nulls in each pulsar. 

Existing studies have established the absence of significant correlations between nulling fraction and fundamental pulsar characteristics including spin period or its temporal derivative \citep{2007MNRAS.377.1383W}. 
PSRs J1649+2533, J1819+1305 and J1916+1023 have similar period and period derivative which mean the positions of the three pulsar on P-$\dot{P}$ are very close. However, This is also true here for the three pulsars. The null/burst distribution, modulation properties, pulse energy distributions for the three pulsars are also quite different. Despite similar
NFs, the positions of PSRs J1752+2359 and J1916+1023 in the P–$\dot{P}$ diagram are not close to each other. 

Drifting and nulling have been detected in hundreds of pulsars. But they were
seen to coexist in only a few pulsars \citep{2017arXiv170605407G}. 
PSR J1916+1023 is a pulsar with both drifting and nulling.
Subpulse drifting is known to be a common phenomenon among pulsars, and its coexistence with nulling has been reported in other pulsars from both FAST observations 
\citep[e.g.,][]{2023RAA.23..104001} and earlier studies with GMRT and Parkes \citep[e.g.,][]{2017APJ.846..109,2020APJ.889..133,2008MNRAS.385.1923R}. 
These results suggest that the simultaneous presence of nulling and drifting may be a relatively common phenomenon among pulsars, although the physical connection between the two remains unclear. 
In our FAST data, the drifting behaviour appears to persist across the emission windows between nulls, with no obvious change in immediately before or after nulls. This suggests that, at least in for the pulsar, the nulling phenomenon may not strongly disrupt the underlying drifting pattern. However, given the limited number of long nulls with sufficient S/N, we cannot rule out more subtle correlations, and higher-sensitivity, longer-duration observations will be needed to fully explore any interaction between the two phenomena.

For these pulsars, the nulling behavior could be associated with the geometrical properties of a drifting “carousel” of sparks \citep{2008MNRAS.385.1923R} 
whose rotation speed is determined by the charge density in the vicinity above the polar cap. The subpulse drift rate could remain unchanged through nulls, indicating these brief nulls are merely interruptions in the drifting emission beam or drop to zero pattern \citep{1978MNRAS.182..711U,2007MNRAS.380..430H}. 
For PSR J1916+1023, the drift rate remains stable through short nulls. However, for the long nulls, it is hard to determine whether the drift rate change or during long nulls since the P3 could slightly change. No nulling–drifting interaction is identified in PSR J1916+1023 \citep{2017arXiv170605407G}.

In summary, our FAST observations of four nulling pulsars have revealed several new aspects of their emission behavior. For PSR J1752+2359, we confirm the quasi-periodic switching reported previously but further show that its so-called ``quasi-null'' or ``RRAT-like'' state is better interpreted as a combination of low-level emission and occasional bright pulses. For PSR J1649+2533, we find evidence for mode changing rather than subpulse drifting, in contrast to earlier low-frequency results. For PSR J1916+1023, both nulling and subpulse drifting are clearly detected, but no direct correlation between the two phenomena is found. Finally, for PSR J1819+1305, our analysis confirms the previously reported modulation. Taken together, these results, when placed in the context of recent large-sample studies \citep[e.g.,][]{2020ApJ...897....8W,2020MNRAS.408..906,2023APJ.948..32,2023RAA.23..104001}
, demonstrate the diverse manifestations of pulsar nulling and related phenomena. They also highlight the power of FAST to uncover subtle emission states and modulation patterns, providing  constraints on pulsar emission physics.

\begin{acknowledgements}

This work was funded by the following grants: Major Science and Technology Program of Xinjiang Uygur Autonomous Region, grant number 2022A03013-4; Zhejiang Provincial Natural Science Foundation, grant number LY23A030001; Natural Science Foundation of Xinjiang Uygur Autonomous Region, grant number 2022D01D85; National Natural Science Foundation of China, grant numbers 12041304.

\end{acknowledgements}

\label{lastpage}

\newpage
\onecolumn

\appendix                  

\section{Single-pulse stacks of Four Pulsars}

\begin{figure}[h]
    \centering
    \includegraphics[width=0.6\textwidth]{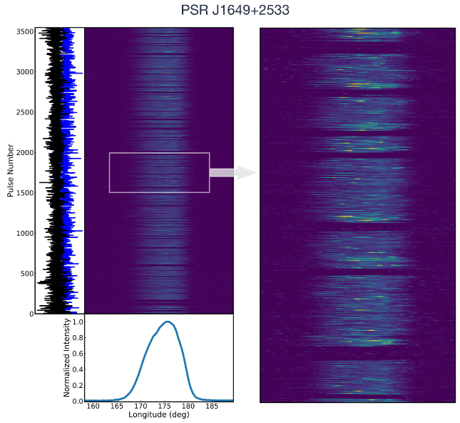}
    \caption{Single-pulse stack of PSR J1649+2533. The blue and black lines of left panel shows the on-pulse and off-pulse energy variations, respectively. The bottom of left panel shows the integrated pulse profiles normalized to the peak intensity. Enlargements for a small part of the whole the pulse stack are shown in the right panel.}
    \label{fg:1649stack}
\end{figure}

\begin{figure}[h]
    \centering
    \includegraphics[width=0.6\textwidth]{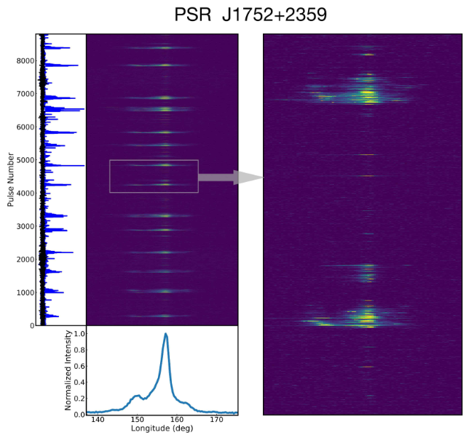}
    \caption{Single-pulse stack of PSR J1752+2359. The blue and black lines of left panel shows the on-pulse and off-pulse energy variations, respectively. The bottom panels show the integrated pulse profiles normalized to the peak intensity.}
    \label{fg:1752stack}
\end{figure}

\begin{figure}[h]
    \centering
    \includegraphics[width=0.6\textwidth]{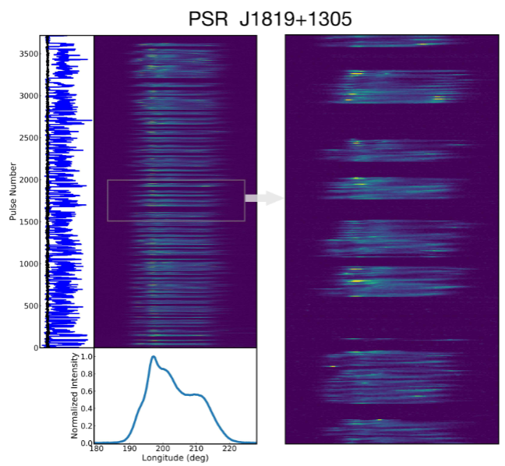}
    \caption{Single-pulse stack of PSR J1819+1305. The blue and black lines of left panel shows the on-pulse and off-pulse energy variations, respectively. The bottom panels show the integrated pulse profiles normalized to the peak intensity.}
    \label{fg:1819gray}
\end{figure}

\begin{figure}[h]
    \centering
    \includegraphics[width=0.4\textwidth]{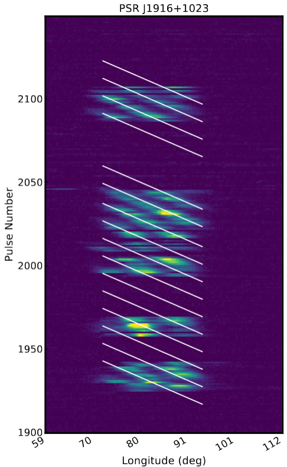}
    \caption{Single-pulse stack of PSR J1916+1023. The white straight lines indicate the linear fitting for each driftband.}
    \label{fg:1916gray}
\end{figure}

\newpage
\onecolumn
\section{Integrate profiles of Four Pulsars}

\begin{figure}[h]
    \centering
    \includegraphics[width=0.9\textwidth]{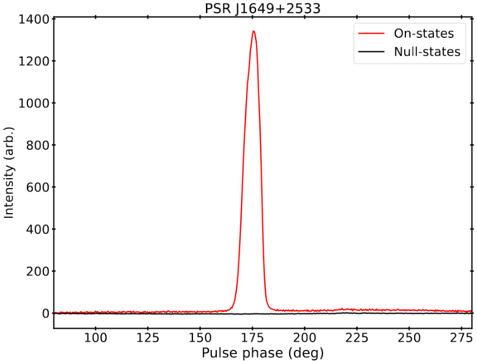}
    \caption{Integrated profile of PSR J1649+2533.}
    \label{fg:1649Intergrate}
\end{figure}

\begin{figure}[h]
    \centering
    \includegraphics[width=0.9\textwidth]{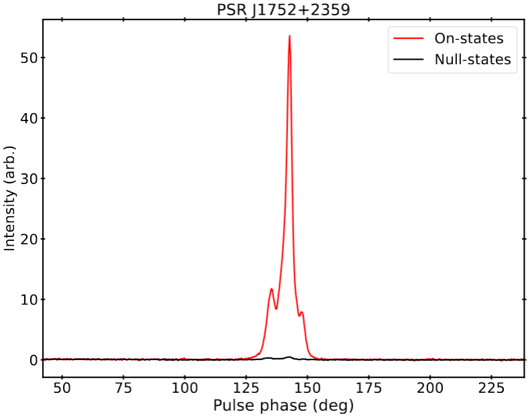}
    \caption{Integrated profile of PSR J1752+2359.}
    \label{fg:1752Intergrate}
\end{figure}

\begin{figure}[h]
    \centering
    \includegraphics[width=0.9\textwidth]{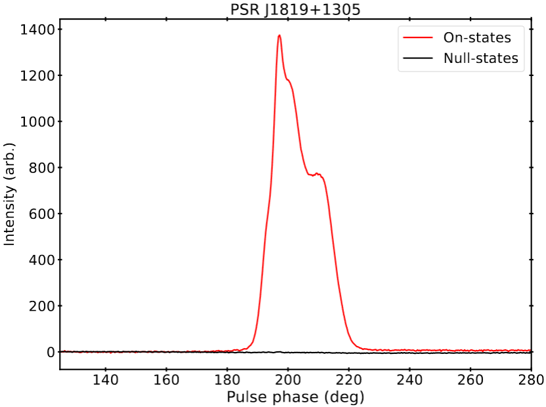}
    \caption{Integrated profile of PSR J1819+1305.}
    \label{fg:1819Intergrate}
\end{figure}

\begin{figure}[h]
    \centering
    \includegraphics[width=0.9\textwidth]{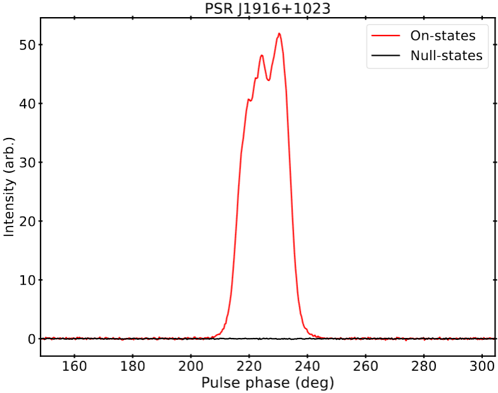}
    \caption{Integrated profile of PSR J1916+1023. }
    \label{fg:1916Intergrate}
\end{figure}

\end{document}